\documentclass[iop,revtex4, apjfonts]{emulateapj}
\usepackage{amssymb,amsmath,mathrsfs,graphicx, url}
\usepackage[colorlinks=true, citecolor=blue, urlcolor=blue, linkcolor=blue]{hyperref}
\usepackage{verbatim}
\usepackage{subfigure}
\usepackage{xcolor}
\tabletypesize{\footnotesize}

\def\h2{$\rm H_2$}

\def\kms{km~s$^{-1}$}

\newcommand{\msun}{M$_{\odot}$}

\newcommand{\halpha}{H$\alpha$}

\newcommand{\hi}{H{\sc I}}

\newcommand{\lsun}{L$_{\odot}$}

\begin{document}

\shortauthors{Weisz et al.}
\title{The Star Formation Histories of Local Group Dwarf Galaxies \sc{II}. Searching For Signatures of Reionization\altaffilmark{*}}

\author{
Daniel R.\ Weisz\altaffilmark{1,2,7},
Andrew E.\ Dolphin\altaffilmark{3}, 
Evan D.\ Skillman\altaffilmark{4},
Jon Holtzman$^{5}$,
Karoline M. Gilbert\altaffilmark{2,6},
Julianne J.\ Dalcanton\altaffilmark{2},
Benjamin F. Williams\altaffilmark{2}
}

\altaffiltext{*}{Based on observations made with the NASA/ESA Hubble Space Telescope, obtained from the Data Archive at the Space Telescope Science Institute, which is operated by the Association of Universities for Research in Astronomy, Inc., under NASA constract NAS 5-26555.}
\altaffiltext{1}{Department of Astronomy, University of California at Santa Cruz, 1156 High Street, Santa Cruz, CA, 95064 USA; drw@ucsc.edu}
\altaffiltext{2}{Department of Astronomy, University of Washington, Box 351580, Seattle, WA 98195, USA}
\altaffiltext{3}{Raytheon Company, 1151 East Hermans Road, Tucson, AZ 85756, USA}
\altaffiltext{4}{Minnesota Institute for Astrophysics, University of Minnesota, 116 Church Street SE, Minneapolis, MN 55455, USA}
\altaffiltext{5}{Department of Astronomy, New Mexico State University, Box 30001, 1320 Frenger St., Las Cruces, NM 88003}
\altaffiltext{6}{Space Telescope Science Institute, 3700 San Martin Drive, Baltimore, MD, 21218, USA}
\altaffiltext{7}{Hubble Fellow}

\begin{abstract}

We search for signatures of reionization in the star formation histories (SFHs) of 38 Local Group dwarf galaxies (10$^4$ $<$ M$_{\star}$ $<$ 10$^9$ M$_{\odot}$). The SFHs are derived from color-magnitude diagrams using archival Hubble Space Telescope/Wide Field Planetary Camera 2 imaging. Only five quenched galaxies (And V, And VI, And XIII, Leo IV, Hercules) are consistent with forming the bulk of their stars before reionization, when full uncertainties are considered. Observations of 13 of the predicted `true fossils' identified by Bovill \& Ricotti show that only two (Hercules and Leo IV) indicate star formation quenched by reionization. However, both are within the virial radius of the Milky Way and evidence of tidal disturbance complicates this interpretation. We argue that the late-time gas capture scenario posited by Ricotti for the low mass, gas-rich, and star-forming fossil candidate Leo T is observationally indistinguishable from simple gas retention. Given the ambiguity between environmental effects and reionization, the best reionization fossil candidates are quenched low mass field galaxies (e.g., KKR 25).

\end{abstract}


\section{Introduction}
\label{intro}

Reionization of the universe is thought to shape the early evolution of low mass galaxies \cite[e.g.,][]{babul1992, efstathiou1992, bullock2000, benson2002, somerville2002, benson2003, kravtsov2004, madau2008, munoz2009, busha2010, simpson2013}.  Preheating from the ultra-violet (UV) background can heat the intergalactic medium (IGM), thereby preventing baryons from falling into the smallest sub-halos.  It can also heat the interstellar medium (ISM) which suppresses star formation within larger mass sub-halos.  The net effect is both to reduce the number of low mass sub-halos that contain stars and to delay or quench star formation in sub-halos that do contain stars \citep[e.g.,][]{gnedin2000, ricotti2002a, ricotti2002b}.  However, because of the rapid decrease in the UV energy density with cosmic time, reionization is only a significant effect at very early times \citep[z $\sim$ 6-14, 12.9 - 13.5 Gyr ago;][]{fan2006}.  Thus, the key signature of reionization will be the suppression of baryons early in the lives of small galaxies.

In principle, signatures of reionization should be detectable within the Local Group (LG).  LG dwarf galaxies (10$^{4}$ $\lesssim$ M$_{\star}$ $\lesssim$ 10$^{9}$ \msun) span the mass range believed to be affected by reionization. Further, their close proximities allow us to resolve individual stars and measure their star formation histories (SFHs), providing a direct way to gauge the impact of reionization on low mass galaxies.   

However, identifying and interpreting signatures of reionization in the LG has proven to be complicated. Most LG dSphs have extended SFHs and/or large metallicity spreads that are incompatible with reionization's early truncation of star formation \citep[e.g.,][]{grebel2004}.  Confronted by observations of extended star formation in dSphs, reionization models have been revised, and now allow for some star formation following reionization (e.g., gas can be preserved via self-shielding or re-accreted gas at later times).  Therefore, ``true fossils'' of reionization are currently viewed as galaxies that formed the bulk of their stars prior to having star formation suppressed or quenched by reionization, \citep[e.g.,][]{ricotti2005, gnedin2006, bovill2009, ricotti2009, bullock2010, bovill2011a, bovill2011b}.

Prospects for identifying fossils have increased with the discovery of faint dwarf galaxies during the SDSS era \citep[e.g.,][]{martin2004, willman2005, belokurov2006, martin2006, zucker2006, zucker2006b, belokurov2007, irwin2007, belokurov2009, belokurov2010, giovanelli2013, rhode2013}.  These extremely low mass (M$_{\star}$ $\lesssim$ 10$^{6}$ \msun) and metal poor ($\langle$[Fe/H]$\rangle$ $\lesssim$ $-$2) systems share many characteristics (e.g., mass, surface brightness) with predicted fossils. Further, several studies have confirmed that many of the faintest galaxies contain stars predominantly older than $\sim$ 11-12 Gyr \citep[e.g.,][]{dejong2008a, sand2009, sand2010, brown2012, okamoto2012, sand2012}, bolstering the possibility that some of the faintest dwarfs are truly fossils \citep[e.g.,][]{ricotti2010}.     

However, the competing influence of environment can complicate true fossil identification. Virtually all of the faintest galaxies known are located within the virial radius of the Milky Way (MW).  Such close proximity to a massive host may enable mechanisms other than reionization to quench early star formation in low mass galaxies \citep[e.g., tidal stripping, tidal stirring, ram pressure;][]{penarrubia2008, lokas2012, gatto2013}. Further, observations show that at least some fraction of the faintest dwarfs display signatures of tidal disturbance \citep[e.g.,][]{martin2008,sand2009, sand2010, blana2012, kirby2013b}, making it challenging to unambiguously identify them as fossils \citep[e.g.,][]{bovill2011b}.  

To minimize environmental considerations, the Local Cosmology from Isolated Dwarfs program (LCID) focused solely on isolated dwarfs in the LG.  With SFHs derived from color-magnitude diagrams (CMDs) that extend below the oldest main sequence turnoff (MSTO), they find that none of the isolated galaxies studied (Cetus, Tucana, LGS~3, Leo~A, IC~1613) are compatible with any reionization-quenching scenario due to their extended SFHs \citep[e.g.,][]{cole2007, monelli2010b, monelli2010c, hidalgo2011, hidalgo2013, skillman2014}.  However, there are subtle caveats about this sample's compatibility with reionization models. IC~1613 (M$_{\star} \sim$ 10$^9$ \msun) is too massive to be significantly affected by reionization. LGS~3, Cetus, and Tucana may have had some past interactions with the MW or M31 \citep[e.g.,][]{lewis2007, fraternali2009, mcconnachie2012}, which could have influenced their early SFHs.   The remaining galaxy, Leo~A, has a suppressed SFH that is unique among known dwarf galaxies \citep{cole2007}, making it an intriguing case study for the early evolution of isolated low mass galaxies.  However, it is challenging to draw broad conclusions from a single system.

To date, there has not been a comprehensive comparison of reionization models and the SFHs of LG dwarf galaxies.  \citet{grebel2004} qualitatively depicted LG dwarf SFHs in the context of reionization.  Since then, the LCID team presented well-constrained SFHs, but for only a handful of galaxies. The SFHs of the faintest dwarfs have typically been studied for single galaxies or in small groups \citep[e.g.,][]{sand2009, sand2010, brown2012, okamoto2012, sand2012, clementini2012, weisz2013b}. The sole largest study of `ultra-faint' dwarfs has insufficiently deep CMDs to resolve the SFHs near the epoch of reionization \citep{dejong2008a}.  

In this paper, we undertake a search of the signatures of reionization in the SFHs of 38 LG dwarf galaxies.  We use the SFHs from \citet{weisz2014a} that were uniformly derived from optical CMDs constructed from archival imaging taken with the Hubble Space Telescope / Wide Field Planetary Camera 2 \citep[HST/WFPC2;][]{holtzman1995}.   In this paper, we have three goals: (1) to provide an empirical assessment of the effects of reionization on the majority of LG dwarfs; (2) to compare our SFHs with predictions froŒm the models of \citet{bovill2011b}; and (3) to discuss the state of observations and models for unambiguously identifying galaxies that formed before reionization.

This paper is organized as follows.  In \S \ref{sec:obs} we briefly describe our sample and in \S \ref{sec:analysis} we summarize the method of measuring SFHs.  We empirically compare our SFHs relative to the epoch of reionization in \S \ref{sec:empiricalreionization} and evaluate them relative to the models of \citet{bovill2011b} in \S \ref{sec:reionizationmodels}.  We discuss our findings and comment on the current state of observations and models in \S \ref{sec:discussion}.  Our findings are summarized in \S \ref{sec:summary}. Throughout this paper, the conversion between age and redshift assumes the Planck cosmology as detailed in \citet{planck2013}.  

This work is part of a series that leverages our uniformly derived SFHs to better understanding the evolution of low mass galaxies and of the LG.  In the first paper, \citet{weisz2014a}, we describe the technical details of the SFH measurements, characterize the SFHs of the LG dwarfs, and provide SFH data for community use.  In future papesr, we will explore the quenching timescales of LG dSphs and compare our SFHs to state-of-the-art simulations of low mass galaxies. We note that all SFH data used in this paper are available electronically in \citet{weisz2014a} and at \textcolor{blue}{\url{http://people.ucsc.edu/~drweisz}}.

\section{The Data}
\label{sec:obs}

Our sample consists of 38 diverse LG dwarf galaxies that were selected based on the availability of deep archival HST/WFPC2 imaging.  It includes several gas-poor, faint dwarfs from both the MW and M31 sub-groups (e.g., Cannes~Vennetici {\sc I}, Hercules, Leo {\sc IV}, And {\sc XI}, And {\sc XII}, And {\sc XIII}) whose stellar masses are $\lesssim$ 10$^5$ \msun.   At the other extreme, it contains more massive, gas-rich, and star-forming galaxies such as IC~1613, Leo~A, and NGC~6822 that have M$_{\star}$ $\gtrsim$ 10$^7$ \msun\ and are located in the `field' of the LG.  The sample also hosts unusual objects such as transition dwarfs, which have \hi\ but no evidence of on-going star formation (i.e., no detectable \halpha), dwarf ellipticals, and the enigmatic Leo~T, which has a low stellar mass ($\sim$ 10$^5$ \msun), evidence for recent star-formation, and a high baryonic gas fraction \citep[e.g.,][]{irwin2007, ryanweber2008}.  Overall, our sample includes 21 dwarf spheroidals (dSphs), 4 dwarf ellipticals (dEs; including compact elliptical M32), 8 dwarf irregulars (dIs), and 5 transition dwarfs (dTrans), making it representative of the entire LG dwarf population.  Observational properties of our galaxy sample are listed in Table \ref{tab:alan_data}.

All galaxies in our sample were observed with HST/WFPC2.  Fluxes of point like sources were measured using \texttt{HSTPHOT}, a CMD fitting package designed specifically for the under-sampled point spread function of WFPC2 \citep{dolphin2000a}, as part of  the Local Group Stellar Populations Archive pipeline \footnote{http://astronomy.nmsu.edu/logphot} \citep{holtzman2006}.  For each CMD, we ran $\sim$  10$^5$ artificial star tests, i.e., we inserted stars of a known magnitude into our images and recovered their photometry in an identical manner to the photometry of real stars, to characterize the completeness and observational biases.  A more detailed description of sample selection and data reductions are presented in \citet{weisz2014a}.

\begin{deluxetable*}{ccccccccc}
\tablecaption{Observational Properties of our Local Group Dwarf Sample}
\tablehead{
\colhead{Galaxy} &
\colhead{Morphological} &
\colhead{M$_{\mathrm{V}}$} &
\colhead{M$_{\star}$} &
\colhead{M$_{\mathrm{H{\sc I}}}$} &
\colhead{D$_{\mathrm{Host}}$} &
\colhead{r$_{\mathrm{h}}$}  &
\colhead{$\tau_{70}$} &
\colhead{Fossil?}  \\
\colhead{Name} &
\colhead{Type} &
\colhead{} &
\colhead{(10$^{6}$ \msun)} &
\colhead{(10$^{6}$ \msun)} &
\colhead{(kpc)} &
\colhead{(pc)}  &
\colhead{log(yr)} &
\colhead{}  \\
\colhead{(1)} &
\colhead{(2)} &
\colhead{(3)} &
\colhead{(4)} &
\colhead{(5)} &
\colhead{(6)} &
\colhead{(7)}  &
\colhead{(8)} &
\colhead{(9)}  \\
}
\startdata 
Andromeda {\sc I} & dSph & -11.7 & 3.9 & 0.0 & 58 & 672 &  9.88$^{+0.02,0.19}_{-0.0,0.01}$ & RG05 \\
Andromeda {\sc II} & dSph & -12.4 & 7.6 & 0.0 & 184 & 1176 & 9.79$^{+0.05,0.27}_{-0.0,0.0}$ & RG05 \\
Andromeda {\sc III} & dSph & -10.0 & 0.83 & 0.0 & 75 & 479 & 9.94$^{+0.05,0.16}_{-0.01,0.08}$ & RG05, BR11a  \\
Andromeda {\sc V} & dSph & -9.1 & 0.39 & 0.0 & 110 & 315 & 10.0$^{+0.0,0.11}_{-0.02,0.12}$ & \\
Andromeda {\sc VI} & dSph & -11.3 & 2.8 & 0.0 & 269 & 524 & 9.93$^{+0.03,0.18}_{-0.01,0.07}$ & RG05 \\
Andromeda {\sc XI} & dSph & -6.9 & 0.049 & 0.0 & 104 & 157 & 10.11$^{+0.0,0.0}_{-0.08,0.2}$ & BR11a \\
Andromeda {\sc XII} & dSph & -6.4 & 0.031 & 0.0 & 133 & 304 & 9.79$^{+0.11,0.21}_{-0.02,0.07}$ & BR11a\\
Andromeda {\sc XIII} & dSph & -6.7 & 0.041 & 0.0 & 180 & 207 & 10.11$^{+0.0,0.0}_{-0.01,0.26}$& BR11a \\
Carina & dSph & -9.1 & 0.38 & 0.0 & 107 & 250 & 9.56$^{+0.09,0.12}_{-0.01,0.02}$ &\\
Canes~Venatici {\sc I} & dSph & -8.6 & 0.23 & 0.0 & 218 & 564 & 9.97$^{+0.08,0.13}_{-0.01,0.07}$ & BR11a \\
Canes~Venatici {\sc II} & dSph & -4.9 & 0.0079 & 0.0 & 161 & 74 & 10.0$^{+0.01,0.06}_{-0.04,0.05}$ & BR11a \\
DDO~210 & dTrans & -10.6 & 1.6 & 4.1 & 1066 & 458 & 10.11$^{+0.0,0.0}_{-0.15,0.21}$ &\\
Draco & dSph & -8.8 & 0.29 & 0.0 & 76 & 221& 10.03$^{+0.03,0.07}_{-0.0,0.06}$ & RG05, BR11a \\
Fornax & dSph & -13.4 & 20.0 & 0.0 & 149 & 710 & 9.67$^{+0.02,0.09}_{-0.04,0.06}$ &\\
Hercules & dSph & -6.6 & 0.037 & 0.0 & 126 & 330 & 10.11$^{+0.0,0.01}_{-0.06,0.11}$ & BR11a \\
IC~10 & dIrr & -15.0 & 86.0 & 50.0 & 252 & 612 & 9.36$^{+0.02,0.27}_{-0.02,0.02}$ &\\
IC~1613 & dIrr & -15.2 & 100.0 & 65.0 & 520 & 1496 & 9.4$^{+0.0,0.41}_{-0.02,0.02}$ & \\
Leo~A & dIrr & -12.1 & 6.0 & 11.0 & 803 & 499 & 9.3$^{+0.0,0.08}_{-0.02,0.09}$ &\\
Leo~{\sc I} & dSph & -12.0 & 5.5 & 0.0 & 258 & 251 & 9.45$^{+0.04,0.07}_{-0.02,0.02}$ & \\
Leo~{\sc II} & dSph & -9.8 & 0.74 & 0.0 & 236 & 176 & 9.83$^{+0.01,0.06}_{-0.0,0.0}$ &  \\
Leo~{\sc IV} & dSph & -5.8 & 0.019 & 0.0 & 155 & 206 & 10.07$^{+0.0,0.04}_{-0.19,0.19}$ & BR11a \\
Leo~T & dTrans & -8.0 & 0.14 & 0.28 & 422 & 120 & 9.81$^{+0.01,0.05}_{-0.15,0.16}$ & BR11a \\
LGS~3 & dTrans & -10.1 & 0.96 & 0.38 & 269 & 470 & 10.01$^{+0.0,0.02}_{-0.13,0.27}$ & \\
M32 & dE & -16.4 & 320.0 & 0.0 & 23 & 110 &  9.79$^{+0.04,0.31}_{-0.01,0.07}$ & \\
NGC~147 & dE & -14.6 & 62.0 & 0.0 & 142 & 623 & 9.76$^{+0.01,0.23}_{-0.0,0.18}$ &\\
NGC~185 & dE & -14.8 & 68.0 & 0.11 & 187 & 42 & 9.92$^{+0.02,0.08}_{-0.02,0.13}$ &\\
NGC~205 & dE & -16.5 & 330.0 & 0.4 & 42 & 590 & 9.9$^{+0.01,0.09}_{-0.04,0.14}$ &\\
NGC~6822 & dIrr & -15.2 & 100.0 & 130.0 & 452 & 354 & 9.36$^{+0.0,0.42}_{-0.02,0.02}$ & \\
PegDIG & dTrans & -12.2 & 6.61 & 5.9 & 474 & 562 & 9.36$^{+0.0,0.75}_{-0.06,0.08}$ & \\
Phoenix & dTrans & -9.9 & 0.77 & 0.12 & 415 & 454 & 9.86$^{+0.02,0.04}_{-0.03,0.07}$ & RG05, BR11a\\
SagDIG & dIrr & -11.5 & 3.5 & 8.8 & 1059 & 282 & 9.67$^{+0.02,0.43}_{-0.15,0.49}$ &\\
Sagittarius & dSph & -13.5 & 21.0 & 0.0 & 18 & 2587 & 9.6$^{+0.01,0.19}_{-0.02,0.02}$ &\\
Sculptor & dSph & -11.1 & 2.3 & 0.0 & 89 & 283 & 10.06$^{+0.0,0.05}_{-0.0,0.0}$ & RG05\\
Sex~A & dIrr & -14.3 & 44.0 & 77.0 & 1435 & 1029 & 9.81$^{+0.02,0.3}_{-0.16,0.51}$ &\\
Sex~B & dIrr & -14.4 & 52.0 & 51.0 & 1430 & 440 & 9.9$^{+0.1,0.2}_{-0.02,0.06}$ &\\
Tucana & dSph & -9.5 & 0.56 & 0.0 & 882 & 284 & 9.84$^{+0.0,0.19}_{-0.01,0.03}$ & RG05, BR11a\\
Ursa~Minor & dSph & -8.8 & 0.29 & 0.0 & 78 & 181 & 9.98$^{+0.03,0.12}_{-0.01,0.08}$ & RG05, BR11a\\
WLM & dIrr & -14.2 & 43.0 & 61.0 & 836 & 2111 & 9.28$^{+0.0,0.08}_{-0.02,0.02}$ &
\enddata
\tablecomments{Physical properties for galaxies in our sample taken from \citet{mcconnachie2012}.  The stellar mass listed in column (4) is computed from the integrated V-band luminosity and assumes \msun/\lsun $=$ 1.  Column (9) indicates the epoch at which 70\% of the stellar mass formed, including the random uncertainties (first error term) and the total uncertainty (second error term).  Both reflect their respective 68\% confidence intervals.  In column (9), we list galaxies in our sample that were identified as fossils by \citet{ricotti2005} (RG) or \citet{bovill2011a} (BR11a).  This table is available electronically at: \textcolor{blue}{\url{http://people.ucsc.edu/~drweisz}}.} 
\label{tab:alan_data}
\end{deluxetable*}

\section{Measuring the Star Formation Histories}
\label{sec:analysis}

We have measured the SFHs for galaxies in our sample using the software package \texttt{MATCH} \citep{dolphin2002}.  Here we briefly summarize the application of this code to our dataset.  Full details are available in \citet{weisz2014a}.

\texttt{MATCH} constructs a set of simple stellar population (SSP) CMDs using input parameters such as a stellar initial mass function (IMF), binary fraction, fixed bin sizes in age, metallicity, color and magnitude, and a given stellar evolution library.  The code combines sets of SSPs into a composite synthetic CMD, which it then convolves with observational biases (e.g., completeness, color and magnitude offsets) as measured from artificial star tests.  \texttt{MATCH} then uses a Poisson likelihood statistic to compute the probability of the observed CMD given the synthetic CMD.  The SFH that produces the best match is deemed the most likely SFH of the observed populations. 

For each of the SFHs presented in this paper, we chose a Kroupa IMF \citep{kroupa2001} with a mass range of 0.15 to 120 \msun\ and a binary fraction of 0.35, where the mass ratio was drawn from a uniform distribution.  We used the solar-scaled Padova set of stellar libraries from \citet{girardi2010}; it is currently the only set of models that cover the entire age, mass, and metallicity ranges needed to systematically analyze such a diverse collection of galaxies. For each galaxy, the code searched a range of metallicities from [M/H] = $-$2.3 to 0.1, with resolution of 0.1 dex.  We required  the metallicity to monotonically increase with time, which is a necessary requirement for accurate SFHs of CMDs that do not reach the oldest MSTO.  We have verified that this requirement does not alter the SFHs of galaxies with deeper CMDs.  To capture possible non-monotonic variations in metallicity, we also included metallicity dispersion as a parameter with a 1-$\sigma$ value of 0.15 dex.  Finally, we modeled intervening MW foreground populations using the models presented in \citet{dejong2010b}.  

For each SFH we quantified both random and systematic sources of uncertainty.  The random uncertainties arise from a finite number of stars on the observed CMD.  The random uncertainties were quantified using a hybrid Monte Carlo technique \citep{duane1987}, using the implementation described in \citet{dolphin2013}.  This approach provides a sample of 10$^4$ SFHs with a density proportional to the probability density of SFH space, allowing us to define the various confidence intervals around the most likely SFH.  Systematic uncertainties are due to intrinsic uncertainties in the underlying stellar evolution models. We quantify these uncertainties using 50 Monte Carlo realizations that are designed to capture variations in the SFHs due to the choice in stellar model. Details of this technique and its application to this sample are available in \citet{dolphin2012} and \citet{weisz2014a}.  Throughout this paper, the quoted uncertainties represent the 68\% confidence intervals around the best fit SFH.

\section{Results}
\label{sec:reionization}

\subsection{Ancient Star Formation in Local Group Dwarfs}
\label{sec:empiricalreionization}

\begin{figure*}[th!]
\begin{center}
\plotone{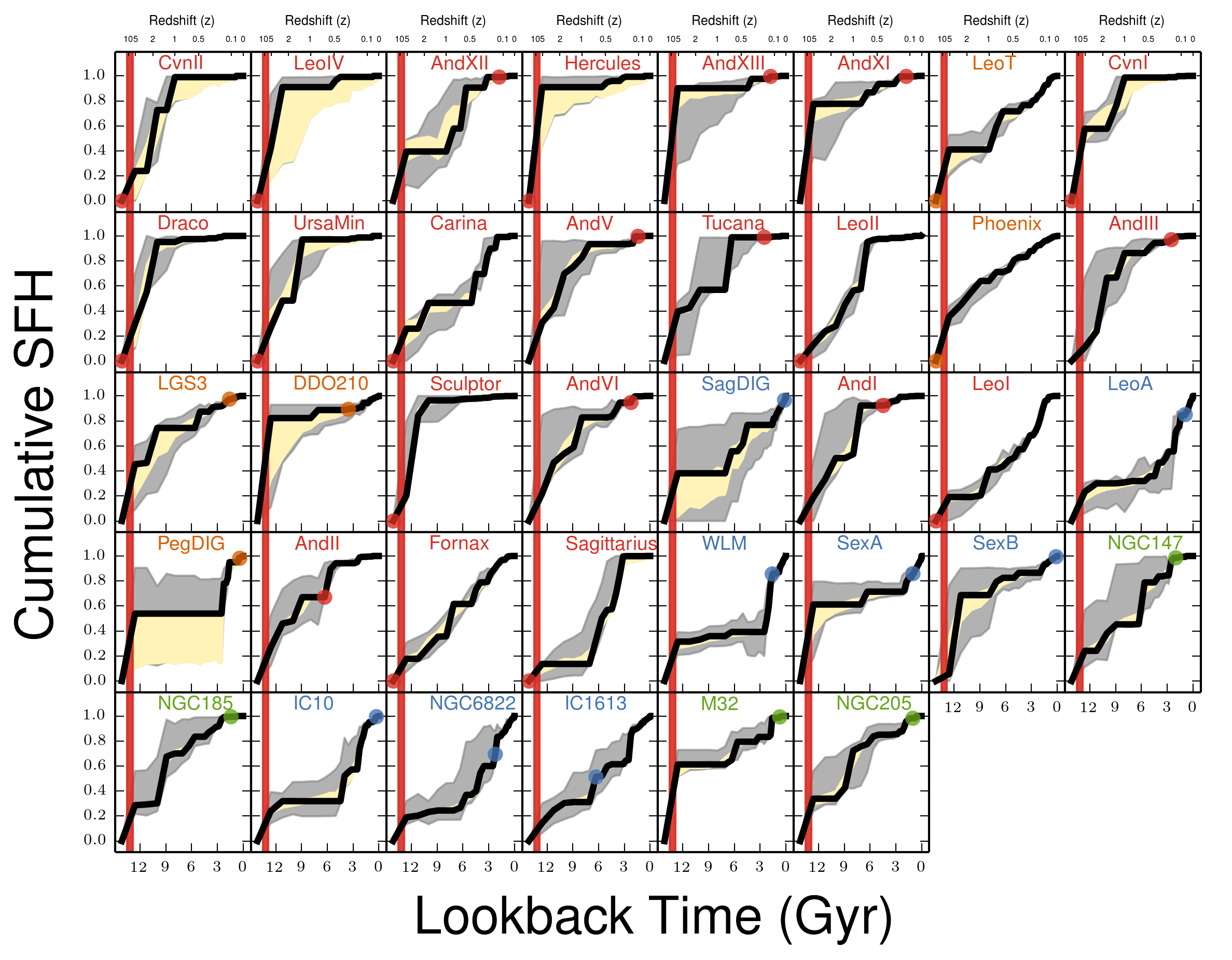}
\caption{The cumulative SFHs, i.e., the fraction of stars formed prior to a given epoch, of the LG dwarfs in our sample, sorted by increasing luminosity.  The solid black lines are the best fit SFHs, the yelloww and grey error envelopes represent the 68\% confidence interval around the best fit for the random and total (random and systematic) uncertainties, respectively.  The colored dots in each panels are the approximate main sequence turnoff age of the CMD used to measure the SFH.  Ages younger than this are considered to be the most secure. The points and text color-coded by morphological type (dSphs, red; dIrrs, blue; dTrans, orange; dEs, green).  The red bands indicate the epoch of re-ionization \citep[z$\sim$ 6-14;][]{fan2006}.  The majority of LG dwarfs did not form the bulk of their stellar mass prior to reionization.  From our sample, only five appear to be good candidates for pre-reionization fossils: And {\sc V}, And {\sc VI}, And {\sc XIII}, Hercules, and Leo {\sc IV}.}
\label{fig:reionization}
\end{center}
\end{figure*}

We begin by comparing features in our SFHs to the epoch of reionization.  In Figure \ref{fig:reionization}, we plot the cumulative SFHs, i.e., the fraction of the total stellar mass formed prior to a given epoch, for our entire sample in order of increasing luminosity.  We also over-plot the commonly accepted range of reionization epochs \citep[z $\sim$ 6-14, 12.9 - 13.5 Gyr ago;][]{fan2006}.  In each panel, we have also indicated the approximate oldest main sequence turnoff (MSTO) age available on the CMD.  The SFH younger than the MSTO are generally the most secure, as they rely on the properties of well-understood MS stars, while SFHs that rely primarily on harder to model evolved stars are less reliable \citep[e.g.,][]{gallart2005}.

This figure shows many of the expected trends.  The dIrrs and dEs show continuous star formation over their lifetimes.  Due to their large masses, these galaxies are not expected to be significantly affected by reionization.  The more luminous MW dSphs, e.g., Draco, Sculptor, Fornax, show star formation that extends past the epoch of reionization, which is not consistent with quenched star formation around z $\sim$ 6, as first pointed out by \citet{grebel2004}.  

In contrast, fainter dSphs are more likely to have stars predominantly older than $\sim$ 10-12 Gyr, which is consistent with the majority of known recently discovered dwarfs \citep[e.g.,][]{dejong2008a, sand2009, sand2010, brown2012, okamoto2012, sand2012}.  When uncertainties are considered, our sample has seven galaxies (Hercules, Leo {\sc IV}, And {\sc XIII}, And {\sc V}, DDO~210, Sculptor, And {\sc VI}) that appear to have formed stars primarily ($>$ 90\%) before reionization.  Of these Hercules and And {\sc XIII} have best fit SFHs that are clearly older than reionization.  Leo {\sc IV} and DDO~210 are the next best candidates, although DDO~210 is gas-rich, making it less likely to be a fossil. Sculptor and And {\sc VI} are marginally good candidates, although \citet{grebel2004} suggest that the metallicity spread in Sculptor makes it a poor fossil candidate.  Some galaxies such as Sag~DIG and Sex~B have large uncertainties at older times that are consistent with predominately ancient star formation.  However, given the amplitude of the uncertainties, their larger stellar masses, and high baryonic gas fractions, we do not consider them to be strong true fossil candidates. 

Observationally measuring a complete shutdown of star formation and attributing it to reionization is challenging.  Some galaxies exhibit subtle features that could be interpreted as causally related to reionization.  For example, the best fit SFH of LGS~3 exhibits no stellar mass growth from $\sim$ 5-10 Gyr ago.  Taken at face value, this suggests 5 Gyr of no star formation that was resumed because of subsequent gas accretion.  While physically plausible, this effect is also a known artifact in CMD-based SFHs due to anti-correlations between nearby time bins that is inherent to the SFH measurement process, particularly for CMDs that do not include the oldest MSTO \citep[e.g.,][]{harris2001, weisz2011a}.  In these situations, interpretations of the random uncertainties become critical.  To capture the full 68\% confidence interval consider the lower edge of the uncertainty envelope at the oldest age of interest and the upper edge of the uncertainty envelope at the younger age of interest.  This range reflect the full 68\% confidence interval.  

In the case LGS~3, the best fit SFH indicates no star formation from $\sim$ 5-10 Gyr ago.  However, over this interval the random uncertainties (yellow envelope) have cumulative fractions ranging from $\sim$ 0.5 as the lower bound at 10 Gyr to $\sim$ 0.8 as the upper bound at $\sim$ 5 Gyr ago.  This range indicates that the 68\% confidence interval is consistent with LGS~3 having formed $\sim$ 30\% of its total stellar mass over these 5 Gyr.  In contrast, Carina, has small random uncertainties over its periods of zero mass growth, which suggests these are more likely to be true periods of no star formation, as has been suggested previously in the literature  \citep[e.g.,][]{smeckerhane1994, hurleykeller1998}.  Typically, when uncertainties are taken into account, there are only a few cases in which there are extended periods of quiescence followed by star formation.

Dwarfs with significantly delayed SFHs are also interesting in the context of reionization.  Strong enough UV preheating may be able to delay star formation until well after reionization ends. Leo~A is a prime example of delayed star formation as it formed $\sim$ 80\% of its stellar mass after z$\sim$2 (10 Gyr ago).  Because of its modest total stellar mass, Leo~A ($\sim$ 10$^7$ \msun) is not predicted to have been affected by reionization, but the extreme suppression of its SFH is clearly intriguing \citep[see][for a detailed discussion]{cole2007}.  We note that no LG dwarfs have had star formation completely suppressed until after reionization, i.e., CMD analysis and/or RR Lyrae observations show that all LG dwarfs had star formation at early times \citep[e.g.,][and references therein]{mateo1998, tolstoy2009, weisz2014a}

\begin{figure*}[th!]
\begin{center}
\plotone{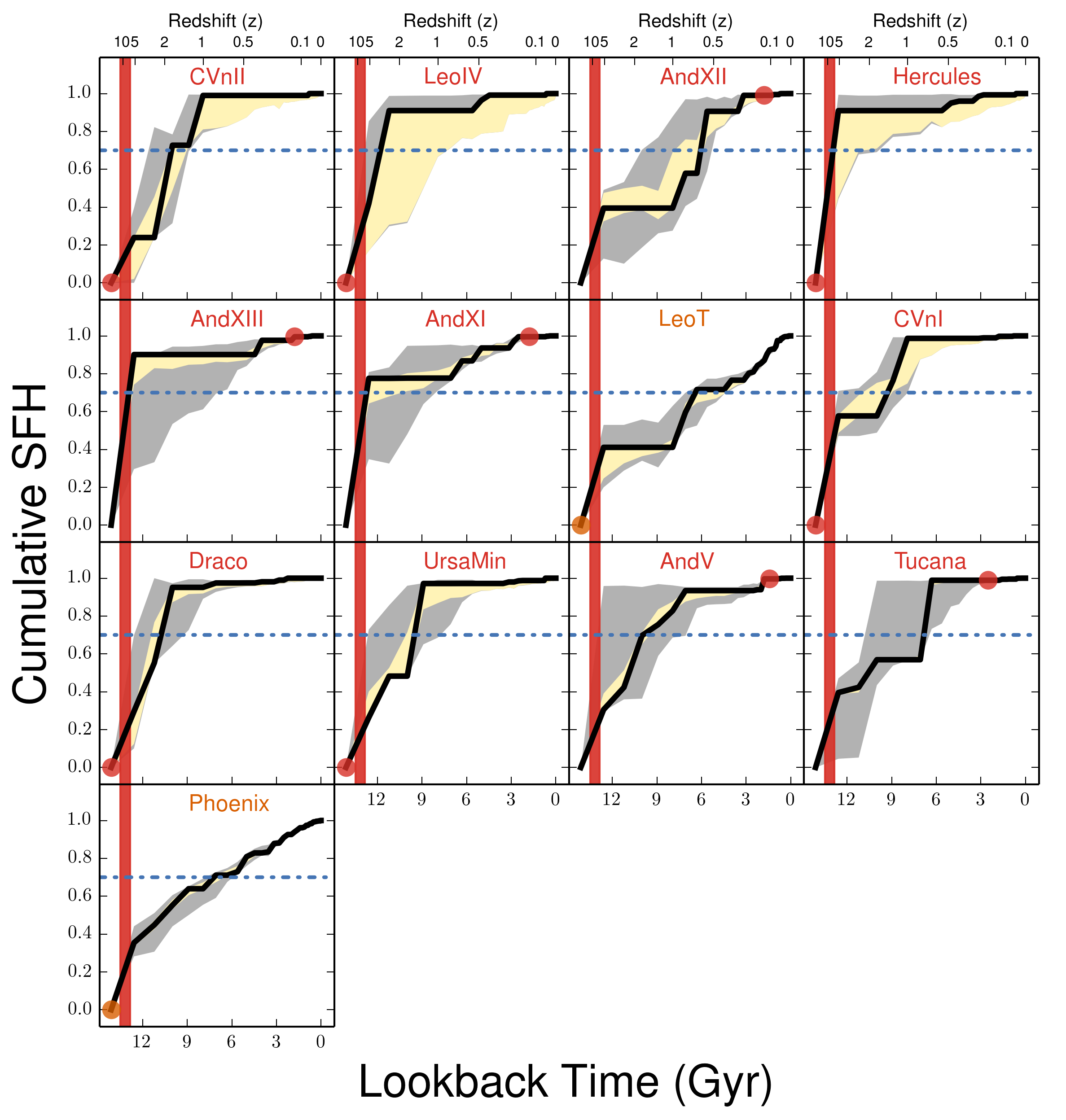}
\caption{Same as in Figure \ref{fig:reionization} only for the 13 `true fossils' identified by \citet{bovill2011b}. The dot-dashed line indicates when 70\% of the stellar mass formed.  As discussed in \S \ref{sec:reionizationmodels}, Hercules and Leo {\sc IV} are the best fossil candidates.  However, their location within the virial radius of the MW and other evidence of tidal disturbance complicates this interpretation.}
\label{fig:reionization2}
\end{center}
\end{figure*}

\subsection{Local Group Star Formation Histories and the Ricotti \& Bovill Models of Reionization}
\label{sec:reionizationmodels}

Among the most directly testable simulations of reionization in the LG are from \citet{bovill2011b}.  These models are the latest in a series that combine high-resolution hydrodynamic and Nbody simulations \citep{gnedin2000, ricotti2002a, ricotti2002b, ricotti2005, gnedin2006, bovill2009, bovill2011a}.  The models of gas hydrodynamics and star-formation come from \citet{ricotti2002a, ricotti2002b}, which only consider baryonic processes up throughout the end of reionization.  The Nbody simulations are performed with \texttt{GADGET-2} \citep{springel2005}, and they are used to track the galaxies' trajectories following reionization.  \citet{bovill2011a} have run new Nbody realizations and incorporate updated observational information, such as the existence of recently discovered faint MW satellites. \citet{bovill2011b} explore the results of these new simulations in the context of the LG.

These models posit that reionization quenches star formation in galaxies whose maximum lifetime circular velocity is $\lesssim$ 20 \kms.  They make detailed predictions for the sizes, surface brightnesses, and SFHs of fossils, and suggest that fossils should have formed  $\gtrsim$ 70\% of their total stellar mass prior to reionization \citep[e.g.,][]{ricotti2005, gnedin2006}.  This hybrid modeling approach does not include features such as the impact of stellar feedback effects on dark matter profiles or interactions such as tidal and ram pressure stripping, both of which may play an important role in the early lives of low mass satellites.  Further, the baryonic components of these models are only simulated until the end of reionization, which means subsequent baryonic processes (e.g., star formation) are not included.  \citet{bovill2011b} present a careful discussion of limitations of their models, and for the remainder of this section, we will take the model predictions at face value.  We discuss some of the caveats in \S \ref{sec:discussion}.  

In Figure \ref{fig:reionization2}, we show the SFHs for the 13 LG dwarfs in our sample identified as fossils by \citet{bovill2011b}.  We find that eight galaxies meet the 70\% criteria (inclusive of all uncertainties): Leo {\sc IV}, Hercules, And {\sc V}, And {\sc XI}, And {\sc XIII}, CVn {\sc I}, Draco, and Ursa~Minor.  Leo~T is also identified as a fossil, and we will discuss this special case separately.  Other galaxies in our sample have all had significant star formation after reionization and are therefore unlikely to be fossils.

Of the eight galaxies, not all are strong fossil candidates. Large systematic uncertainties in the SFHs $>$ 10 Gyr ago of And {\sc V}, And {\sc XI}, and And {\sc XIII} prohibit us from drawing any definitive conclusions.  Of the remaining galaxies, Ursa~Minor and Draco have primarily ancient (8-10 Gyr) populations, but not as old as would be expected from true relics of reionization (i.e., $\gtrsim$ 12 Gyr ago).  Thus, they are not unambiguous candidates for true fossils, which was first suggested by \citet{grebel2004}.
  
Two galaxies, Hercules and Leo {\sc IV}, seem to be the best candidate fossils.  In addition to our SFHs, several other studies have found these galaxies to be predominantly composed of stars $>$ 11-12 Gyr old \citep[e.g.,][]{dejong2008a, sand2009, sand2010, brown2012, okamoto2012}, and they have other properties (e.g., surface brightness) consistent with simulated fossils in \citet{bovill2011b}.  However, there have also been suggestions that these galaxies are tidally disturbed \citep[e.g.,][]{belokurov2007, sand2009, dejong2010a, sand2010, blana2012}, which may complicate their candidacy as pristine fossils.

\section{Discussion}
\label{sec:discussion}

In principle, identifying a fossil is a straightforward problem: find low mass galaxies with SFHs that truncate around the epoch of reionization.  However, there are several competing environmental processes that can also quench star formation in low mass galaxies at early times.  Thus, establishing a causal relationship between reionization and quenching is quite challenging.  In this section, we review some of the mechanisms that can compete with reionization to quench star formation, and discuss future prospects for identifying true fossil candidates.

\subsection{Environmental Quenching Mechanisms}

From our initial sample of 38 galaxies, we have identified two strong fossil candidates, Hercules and Leo {\sc IV}.  However, both galaxies are currently located well within the virial radius of the MW.  Because of their close proximities to the MW, several processes other than reionization could have quenched star formation.

One such mechanism is ram pressure stripping. \citet{gatto2013} show that the hot halo of the MW can efficiently strip gas from dwarfs causing abrupt quenching of their star formation at distances as large as $\sim$ 90 kpc.  Their model predictions are consistent several observational trends including quenching timescales in Carina and Sextans and the spatial distribution of gas-rich and gas-poor LG galaxies \citep[e.g.,][]{grcevich2009}. Additionally, ram pressure provides a clear explanation for the recent quenching of star formation in Leo {\sc I}, which underwent a pericentric passage of the MW $\sim$ 2 Gyr ago at a distance of $\sim$ 90 kpc \citep{sohn2013, boylankolchin2013} and stopped forming stars $\lesssim$ 1 Gyr later \citep[e.g.,][]{weisz2014a}.  Given the efficiency of ram pressure stripping, even at large distance from the MW, it is plausible, and perhaps likely, that ram pressure stripping was a dominant quenching mechanism in many MW dSphs.  However, given the ancient quenching timescales and lack of gas in Hercules and Leo {\sc IV} is it not possible to empirically assess the influence of ram pressure stripping on these particular systems.

Another possible quenching process is `tidal stirring' \citep[e.g.,][]{mayer2001, mayer2006}.  \citet{lokas2012} demonstrate that a combination of tides and ram pressure can transform a gas-rich dwarf into a present day low luminosity dSph.  However, this process may take several Gyr, which does not appear to be compatible with the star formation and quenching timescales of Hercules and Leo {\sc IV}.  

Baryonic physics may also enhance the importance of environment as a quenching mechanism.  For example, several simulations have shown stellar feedback can create cored dark matter profiles in dwarf galaxies \citep[e.g.,][]{governato2010, zolotov2012}.  Cores can lead to weaker effective potentials, making a low mass galaxy more susceptible to tidal and ram pressure effects.  However, whether galaxies with masses as low as Hercules and Leo {\sc IV} have cores or can even create cores remains unclear \citep[e.g.,][and references therein]{penarrubia2012, walker2013}.  Beyond core creation, stellar feedback also heats gas and expels it into the outer halos of low mass galaxies \citep[e.g.,][]{dekel1986}, where it is more susceptible to ram pressure and tidal stripping effects.

Overall, there is some evidence for environmental influence on the lowest luminosity galaxies. Of our best fossil candidates, Hercules appears to have stream-like elongation, and there is debate over whether Leo {\sc IV} and Leo {\sc V} are a paired system \citep[e.g.,][]{martin2008, sand2009, sand2010, dejong2010a, sand2012, blana2012}. Similarly, Segue {\sc I}, Segue {\sc II}, and  Willman {\sc I}, have metallicities higher than what is expected from the luminosity-metallicity relationship, which suggest that these are tidally stripped remanents of more massive progenitors \citep[e.g.,][]{belokurov2009, simon2011, willman2011, vargas2013, kirby2013b}.  While these features may have arisen post-reionization, it is currently not possible to unambiguously distinguish between environmental and reionization quenching processes in these galaxies.

\subsection{Reionization and Late Time Gas Accretion}
\label{sec:gasaccretion}

\citet{bovill2011b} identified Leo~T (M$_{\star} \sim$ 10$^6$ \msun) as a fossil candidate, despite its high baryonic gas fraction (f$_{\mathrm{gas}} \sim$ 0.8) and evidence of significant star formation after reionization \citep[e.g.,][]{dejong2008b, ryanweber2008, clementini2012, weisz2012b}.  \citet{ricotti2009} suggests that Leo~T experienced late time gas capture, enabling it to reignite star formation after reionization.  Our SFHs are consistent with a scenario in which Leo~T had no star formation from z$\sim$1-5 ($\sim$ 8-12 Gyr ago).  However, the uncertainties are also consistent with low to modest star-formation over this interval.    This evidence for little to no star-formation only suggests that there was not enough dense enough gas for star formation to proceed.  For example, gas heated by stellar feedback could have resided in Leo~T's halo for an extended period, before Leo~T's extremely shallow potential well finally caused it to re-condense. Thus,  even with well-constrained SFHs in hand,  it remains unclear how to distinguish a late time gas-capture scenario from one in which Leo~T has simply been depleting the same gas reservoir over most of its lifetime.  More detailed modeling and unique predictions for these scenarios are needed to help decipher the observations.

\subsection{The Future of Fossils Identification}

Beyond fossil galaxies, \citet{bovill2011b} make several predictions for signatures of reionization in the nearby universe.  Among these are (1) the presence of `ghost halos' in low mass galaxies, i.e., extremely diffuse, low surface brightness, purely ancient stellar halos that surround isolated dwarf galaxies; (2) gas-rich galaxies in the voids, which may or may not contain stars; and (3) the presence of undetected quenched galaxies outside a massive galaxy's virial radius.  

However, these first two predictions are of limited utility in observationally identifying fossil dwarfs, since both have other, equally plausible physical explanations.  For example, isolated galaxies near the edge of the LG and in the field ubiquitously contain ancient, diffuse halos \citep[$>$ 12 Gyr; e.g.,][and references therein]{dalcanton2009, dacosta2010, hidalgo2013}, and their existence can be explained by mechanisms other than reionization \citep[e.g., stellar feedback, outside-in star-formation; e.g.,][]{stinson2009, hidalgo2013}.  Similarly, as discussed in \S \ref{sec:gasaccretion} low mass and gas-rich galaxies located isolation such as Leo~T and Leo~P \citep{giovanelli2013, mcquinn2013, rhode2013, skillman2013} appear to have continuous star-formation throughout their lifetimes, which can plausibly be explained by simple gas retention and consumption.

The third prediction, however, holds particular promise for true fossil identification. In an isolated context, there are few mechanisms aside from cosmic reionization that appear capable of completely quenching star formation at early times.  However, quenched field galaxies are exceedingly rare \citep{geha2012}.  Within a few Mpc of the LG, there are only a few known examples.  Cetus and Tucana are quenched galaxies located near the periphery of the LG, but interactions with the Milky Way at early times cannot be ruled out \citep[e.g.,][]{lewis2007, fraternali2009}.  Outside the LG, KKR~25 is perhaps the best candidate for an isolated reionization fossil \citep[e.g.,][]{bovill2011b, makarov2012}.  It is an isolated and quenched field galaxy (D$\sim$1.9 Mpc), that appears to only host an RGB, with few, if any, AGB stars., i.e., little intermediate age star formation.  However, its CMD only extends as deep as the red clump, making its early time SFH extremely uncertain \citep[e.g.,][]{weisz2011a}.  

The future of finding fossil candidates is promising.  Large area photometric surveys such as Pan-STARRS and LSST \citep{kaiser2002, ivezic2008} should provide the necessary depth and sensitivity to identify fossil candidates outside the virial radius of the Milky Way \citep[e.g.,][]{bullock2010}. However, precise SFH measurements will require the angular resolving power of instruments such as HST\footnote{JWST has the requisite angular resolution.  However, due to small color separation in the near-IR, distinguishing between 11 and 13 Gyr old stars at the oldest MSTO in a mixed population will be extremely challenging.}.  While ground-based telescopes can be used to reach below the oldest MSTO for galaxies within a few hundred kpc \citep[e.g.,][]{okamoto2012}, those at larger distances suffer from extreme crowding effects that can only be overcome with HST-like resolution.  With the pending retirement of HST, high resolution imaging from current (e.g., LBT LINC-Nirvana) and planned (e.g., GMT, TMT, ELT) telescopes will play a critical role in identifying fossils outside the virial radius of the Milky Way.

\section{Summary}
\label{sec:summary}

We have searched for  signatures of reionization in the uniformly measured SFHs of 38 LG dwarf galaxies.  When full uncertainties are considered, five currently quenched galaxies (Hercules, And {\sc XIII}, And {\sc V}, Sculptor, And {\sc VI}) are consistent with having formed the bulk of their stars prior to reionization.  The LG dwarf galaxy models of \citet{bovill2011b} identify 13 galaxies in our sample as likely `true fossil' candidates, i.e., low mass galaxies whose star formation was quenched by reionization.  Of these only two, Hercules and Leo {\sc IV}, appear to be viable fossil candidates.  However, both show some evidence for tidal disturbance and are located within the virial radius of the MW, which complicates the interpretation.  Several other models of LG dwarf galaxy evolution suggest that mechanisms other than reionization (e.g., ram pressure, tidal stripping of stars and/or gas) can quench star formation in dwarf galaxies at early times.  Galaxies with cored dark matter profiles are particularly vulnerable to stripping, due to a weaker effective potential.  However, it remains unclear whether the progenitors of the lowest mass dwarfs have cores and/or were capable of forming cores. 

\citet{bovill2011b} also identified gas-rich, low mass galaxy Leo~T as a fossil candidate, and attribute the presence of gas to late time accretion \citep{ricotti2009}.  We argued that this scenario is observationally indistinguishable from one in which Leo~T simply retained and consumed gas via star formation.  Additionally, \citet{bovill2011b} predict the existence of ancient `ghost halos' around nearby dwarfs and existence of low mass, gas-rich galaxies in the field are further evidence for the impact of reionization on low mass galaxy evolution.  We argued that these predictions are not unique to quenching by reionization.  

The best fossil candidates are low mass (M$_{\star} \lesssim$ 10$^7$ \msun), quenched field galaxies.  There appear to be few other mechanisms capable of quenching star formation in completely isolated systems.  At present, few such galaxies are known.  The only local example of such a galaxy is KKR~25 (D~1.9 Mpc).  However, it is too distant to measure an accurate SFH around the epoch of reionization, even with HST.  

The future for identifying fossil candidates is promising.  Large area photometric surveys such as Pan-STARRS and LSST have the potential to discover quenched dwarfs outside the virial radius of the Milky Way.  However, only telescopes with HST-like resolution will be able to precisely measure their SFHs around the epoch of reionization. 

\section*{Acknowledgements}

The authors would like to thank the anonymous referee for through and insightful comments that helped improve the scope and clarity of this paper.  DRW would like to thank Oleg Gnedin, Sijing Shen, Eric Bell, Mike Boylan-Kolchin, and Charlie Conroy for insightful discussions about reionization and dwarf galaxies, and also Hans-Walter Rix and the MPIA for their hospitality during the assembly of this paper.  Support for DRW and KMG is provided by NASA through Hubble Fellowship grants HST-HF-51331.01 and HST-HF-51273.01, respectively, awarded by the Space Telescope Science Institute. Additional support for this work was provided by NASA through grant number HST AR-9521 from the Space Telescope Science Institute, which is operated by AURA, Inc., under NASA contract NAS5-26555. This research made extensive use of NASA's Astrophysics Data System Bibliographic Services.  In large part, analysis and plots presented in this paper utilized iPython and packages from Astropy, NumPy, SciPy, and Matplotlib \citep[][]{hunter2007, oliphant2007, perez2007, astropy2013}.

\bibliography{lgdap_reionization.astroph.bbl}

\end{document}